\documentstyle[prd,aps,preprint,epsfig]{revtex}

\tighten

\begin{document}
\draft
 
\pagestyle{empty}

\preprint{
\noindent
\today \\
\hfill
\begin{minipage}[t]{3in}
\begin{flushright}
LBNL--53265 \\
July 2003
\end{flushright}
\end{minipage}
}

\title{Production of axial-vector $D_{sJ}$ in $e^+e^-$ annihilation}

\author{
Mahiko Suzuki
}
\address{
Department of Physics and Lawrence Berkeley National Laboratory\\
University of California, Berkeley, California 94720
}


\date{\today}
\maketitle

\begin{abstract}
If one of the recently discovered charmed-strange mesons 
($D_{sJ}(2317)$) is the $0^+$ state of $c\overline{s}$, 
the other ($D_{sJ}(2460)$) is most likely the $1^+$ 
state with $j=\frac{1}{2}$. They could be produced in $e^+e^-$
annihilation at $E_{e^+e^-}^{{\rm cm}}=m_{\Upsilon(4S)}$ either 
by fragmentation from $c\overline{c}$ jets or as decay 
products of the $B$ mesons from $\Upsilon(4S)\to B\overline{B}$. 
If one analyzes the $c\overline{c}$ jet events and the 
$\Upsilon(4S)$ decay events separately, one will have a direct 
test as to whether $D_{sJ}(2460)$ is the $j=\frac{1}{2}$ state 
or not, how much $D_{sJ}(2460)$ is mixed with the $j=\frac{3}{2}$ 
state, and also whether the four-quark interpretation is viable.

\end{abstract}
\pagestyle{plain}
\narrowtext

\setcounter{footnote}{0}
\section{Introduction}

The BaBar Collaboration\cite{BaBar1} discovered a narrow peak 
at 2317 MeV in the invariant mass of $D_s\pi^0$. The decay into 
$D_s\pi^0$ suggests $J^P =0^+$ for this state $D_{sJ}(2317)$ 
since good candidates already exist for $1^-$ and $2^+$. 
The observation of this peak was subsequently confirmed by 
the CLEO\cite{CLEO} and the Belle Collaboration. 

The BaBar data show another sharp peak at 2460 MeV in the 
invariant mass of $D^*_s\pi^0$ ($D_s\gamma\pi^0$). Although 
BaBar was initially cautious in calling it as a resonance, 
CLEO concluded that it is another narrow resonance. 
Absence of the decay $D_{sJ}(2460)\not\to D_s\pi^0$ leads us 
to speculate that $D_{sJ}(2460)$ is one of $J^P=1^+$ states 
and forms with $D_{sJ}(2317)$ a $j=\frac{1}{2}$ multiplet in 
terms of total momentum ${\bf j}={\bf l}+{\bf s}$ of the 
$\overline{s}$ quark in the heavy-light limit of $c\overline{s}$. 
Then the existing $1^+$ resonance $D_{s1}(2536)$\cite{PDG} 
should be assigned to a $j=3/2$ multiplet with the $2^+$ 
candidate $D_{sJ}(2573)$.

 Theorists have worked on spectroscopy of the heavy-light quark
states for long time. Schnitzer\cite{Sch} argued on the basis 
of the $l$-$s$ coupling in a naive two-body potential model 
that for heavy-light mesons, the $j=\frac{3}{2}$ multiplet 
should be lighter than the $j=\frac{1}{2}$ multiplet. It was 
pointed out many years later that $K$ meson resonances seemed 
to show this ``inversion'' of the spin-orbit coupling sign
despite the relative lightness of the 
$s$-quark\cite{Suzuki}.\footnote{       
Since then, however, the measured branching fractions of  
$\tau\to K_1\nu_{\tau}$ had shifted so that the 
inversion argument is no longer supported by 
$\tau$ decay data. Only the $s$-$d$ ratio of $K_1\to 
K^*\pi/\rho K$ may favor the inversion, if at all.}
Then the spin-orbit inversion in the heavy-light system was 
studied systematically by Isgur\cite{Isgur}. More recently 
a detailed computation was presented in the potential
model\cite{Eichten}. However, the potential picture has an 
uncertainty of long-distance physics since the energy scale of 
the heavy-light potential is the reduced mass, {\em i.e.,} the 
light mass, not by the heavy mass. Upon the discovery of 
$D_{sJ}(2317)$, Cahn and Jackson\cite{Cahn} re-examined the 
$c\overline{s}$ states with the potential model. It is fair 
to say that the potential model is inconclusive about which 
of the $j$-multiplets is heavier than the other. By keeping 
the inversion scenario back in mind, many theorists proposed 
the exotic possibility that $D_{sJ}(2317)$ is a four-quark 
state or its mixing\cite{fourquark,Pakvasa}. A cursory 
examination by lattice QCD was also reported\cite{Bali}.  

Another approach to the heavy-light mesons is to treat the 
light component as a chiral-symmetric cloud instead of 
a constituent quark\cite{Burdman}. Bardeen 
{\em et al} made a strong case for noninversion of 
$j=\frac{1}{2}$ and $j=\frac{3}{2}$ with detailed 
calculations\cite{Bardeen}. Among others the observed 
$D_{sJ}$ mass difference, $m(2460)-m(2317)\simeq 
m(1^-)-m(0^-) =$ 143 MeV, is a successful consequence of 
chiral symmetry. As for magnitude of the splitting between 
$j=\frac{1}{2}$ and $\frac{3}{2}$, however, an earlier 
work\cite{Gatto} had found much larger uncertainty than 
Bardeen {\em et al.} did. 
  
While the case for $j=\frac{1}{2}$ appears strong for 
$D_{sJ}(2317)$ and $D_{sJ}(2460)$, an independent experimental 
confirmation is desirable. At  $E_{e^+e^-}^{({\rm cm})}=$ 
the $\Upsilon(4S)$ mass, $D_{sJ}$ can be produced through 
$c\overline{c}$ jets or $B$ decays, Two types of processes
occur with roughly the same rate and can be separated without 
difficulty by event topology and by $B$ decay vertices.
We point out here 
that production rates of the axial-vector $D_{sJ}$ in 
$c\overline{c}\to D_{sJ}X$ and in $\Upsilon(4S)\to
B\overline{B}\to D_{sJ}X$ are sensitive to $j=\frac{1}{2}$ 
vs $\frac{3}{2}$ and therefore that analyzing the $D_{sJ}(2460)$ 
production in $c\overline{c}$ and in $B\overline{B}$ separately 
will give us an additional clue about as to whether 
$D_{sJ}(2460)$ is $j=\frac{1}{2}$ or $\frac{3}{2}$ 
of $c\overline{s}$, or else a four-quark state 
$c\overline{s}q\overline{q}$. The two types of production
occur with roughly the same rate and separable by event
topology and $B$ decay vertices.

\section{$B$ decay}

Production of charmed strange mesons is one of the dominant 
nonleptonic $B$ decay processes. It occurs mainly
through the effective decay operators of tree type:
\begin{eqnarray}
   {\cal O}_1 &=& (\overline{c}s)_{V-A}(\overline{b}c)_{V-A},\;\;\;
       \\ \nonumber 
   {\cal O}_2 &=& (\overline{c}c)_{V-A}(\overline{b}s)_{V-A},
\end{eqnarray}
where colors are contracted within each bracket. Although detailed 
comparison between theory and experiment has not been available
for $B\to D_s$, the similar decays $B\to\overline{D}\pi$, 
$\overline{D}^*\pi$, $\overline{D}\rho$ and so forth that occur 
through $(\overline{u}d)_{V-A}(\overline{b}c)_{V-A}$ and 
$(\overline{u}c)_{V-A}(\overline{b}d)_{V-A}$ have been measured 
with good accuracy\cite{PDG}. These color-favored two-body decays 
agree well with the theoretical values computed in the 
factorization approximation. The factorization is even simpler 
for the inclusive color-favored decays since they are free
from the quark distribution involving the spectator quark.
We therefore proceed by assuming that the color-favored decay 
$B\to D_{sJ}X$ is described by the factorization.\footnote{
This is not true for the color-suppressed decays such as 
$B^0\to\overline{D}^0\pi^0$ and charmonium production.
The factorization-forbidden charmonia of $J^{PC}=0^{++}$ and 
$2^{++}$ are abundantly produced.\cite{Belle2}}
In the factorization limit only $D_{sJ}$ of $0^-$, $1^-$, and 
$1^+$ can be produced. As for the $1^+$ states, we shall see 
below that the $1^+$ state of $j=\frac{1}{2}$ would be produced 
preferentially for $m_s/m_c\ll 1$.  

   In the simple factorization the inclusive decay 
$B\to D_{sJ}X$ is determined by the short-distance quark decay 
of $\overline{b}\to\overline{c}(c\overline{s})$. While inclusive 
production of $0^-$ and $1^-$ in raw data contains the 
contribution from strong and electromagnetic cascade decays 
of higher $D_{sJ}$ states, production of $1^+$ is very likely 
free of such contamination, The reason is that only the radially 
excited $0^-$ and $1^{\pm}$ states below the $DK$ threshold 
are possible sources of cascade decays down to the $1^+$ states
in the factorization. Such excited states have not been seen 
in experiment. They are expected to be above the $DK$ threshold. 
Therefore the $D_{s1}(2460)$ reconstructed in $B$ decay may be 
counted entirely as primary decay products of $B$ meson.
 
The production amplitude for $\overline{b}\to 1^+\overline{c}$ 
is given by
\begin{equation}
  A(\overline{b}\to D_{s1j}\overline{c}) =
        (G_{\mu}/\sqrt{2})V_{cb}^*V_{cs}
        (C_1+C_2/3)f_{Aj}m_{D_{s1j}}
    \epsilon^{\mu}(\overline{v}_b\gamma_{\mu}\gamma_5 v_c), \;\;\;
                       (j=1/2,3/2) \label{amp}
\end{equation} 
where the axial-vector decay constant $f_{Aj}$ is defined
with the normalization $\langle{\bf p}|{\bf p}'\rangle=(2\pi)^3
2E_{\bf p}\delta({\bf p}-{\bf p}')$ by
\begin{equation}
 \langle D_{s1j}(p,\epsilon)|(\overline{c}\gamma^{\mu}\gamma_5s)
   |0\rangle
          = f_{Aj}m_{D_{s1j}}\epsilon^{\mu}. \label{DC}
\end{equation}
We have introduced the additional subscript $j$ for $D_{s1}$ 
to distinguish between two eigenstates of ${\bf j}^2$.
If one evaluates Eq. (\ref{DC}) in the rest frame of $D_{s1j}$ 
by treating $D_{s1j}$ as being made of the $c$-quark and the 
remainder carrying the $\overline{s}$-quark quantum numbers 
(still denoted by $\overline{s}$), the left-hand side is written
in the Pauli spinors as 
\begin{eqnarray}
       \langle D_{s1j}({\bf 0},\epsilon,j)|(\overline{c}
         \mbox{\boldmath$\gamma$}\gamma_5s)|0\rangle 
    = N\Biggl[
   \biggl(\frac{1}{E_s+m_s} &+& \frac{1}{E_c+m_c}\biggr)
   i\chi^{\dagger}_c({\bf p}_s\times\mbox{\boldmath$\sigma$})
               \chi_{\overline{s}}     \nonumber \\ 
    &+&  \biggl(\frac{1}{E_s+m_s}-\frac{1}{E_c+m_c}\biggr)
       \chi^{\dagger}_c{\bf p}_s\chi_{\overline{s}}\Biggr],
                                 \label{current}
\end{eqnarray}
where $N$ is an normalization factor. The spinors form 
$^3P_1$ and $^1P_1$ in the first and the second term, 
respectively, in the right-hand side of Eq. (\ref{current}). 
In the limit of $m_s/m_c\to 0$, the right-hand side 
approaches $\chi^{\dagger}\mbox{\boldmath$\sigma$}
(\mbox{\boldmath$\sigma$}\cdot{\bf p}_s)\chi_{\overline{s}}$
up to an overall constant, which is exactly the 
$j=\frac{1}{2}$ combination of $^3P_1$ and $^1P_1$. 
Therefore, the weak axial-vector current can produce only 
the $j=\frac{1}{2}$ state of $D_{s1}$ in the large $m_c$
limit\cite{Colangelo}. For $m_s/m_c\neq 0$, the weak 
current produces the combination of
\begin{equation}
          |D_{s1\frac{1}{2}}\rangle\cos\alpha
           - |D_{s1\frac{3}{2}}\rangle\sin\alpha, 
\end{equation} 
where
\begin{equation}
  \tan\alpha =\frac{2\sqrt{2}}{3(E_c+m_c)/(E_s+m_s)+1}
                        \label{mixture} 
\end{equation}
in the phase convention of
$|j=\frac{1}{2}\rangle = \sqrt{\frac{1}{3}}|^1P_1\rangle -
\sqrt{\frac{2}{3}}|^3P_1\rangle$ and $|j=\frac{3}{2}\rangle 
= \sqrt{\frac{2}{3}}|^1P_1\rangle +
\sqrt{\frac{1}{3}}|^3P_1\rangle$.  For a nonrelativistic 
binding with $m_s/m_c\simeq \frac{1}{3}$, the mixture of 
$|j=\frac{3}{2}\rangle$ is small ($\tan^2\alpha\simeq 0.09$ 
in probability) and production of $D_{s1\frac{3}{2}}$ is 
almost negligible. To obtain the production rates 
of the mass eigenstates, one needs to know about 
a small mixing between $j=\frac{1}{2}$ and $\frac{3}{2}$ 
in the mass eigenstates:
\begin{eqnarray}
    |D_{s1}(2460)\rangle &=& \cos\theta|D_{s1\frac{1}{2}}\rangle
             -\sin\theta|D_{s1\frac{3}{2}}\rangle, \nonumber \\
    |D_{s1}(2536)\rangle &=& \sin\theta|D_{s1\frac{1}{2}}\rangle
             +\cos\theta|D_{s1\frac{3}{2}}\rangle. 
\end{eqnarray}
Then the ratio of the branching fractions for two mass 
eigenstates is 
\begin{equation}
  \frac{{\cal B}(B\to D_{s1}(2536)X)}{{\cal B}(B\to D_{s1}(2460)X)}
                     =\tan^2(\alpha-\theta). \label{B}   
\end{equation}
In the case that $D_{s1}(2460)$ consists mostly of $j=\frac{1}{2}$,
the mixing angle $\theta$ is $O(m_s/m_c)$ so that the production
rate of $D_{s1}(2536)$ is one order of magnitude smaller than 
that of $D_{s1}(2460)$.  Theoretical estimate of the value 
of $\theta$ is not possible because of unknown long-distance 
effects. We should use Eq. (\ref{B}) to determine the mixing 
angle $\theta$ albeit Eq. (\ref{mixture}) has some 
model dependence.

If the value of the decay constant $f_{Aj}$ is given for 
$j=\frac{1}{2}$, we can compute the branching fraction of the 
inclusive $B\to D_{s1\frac{1}{2}}X$ decay. The authors in 
\cite{Colangelo} gave one estimate, which corresponds to 
$f_A(\equiv f_{A\frac{1}{2}}) \simeq 200$MeV for $m_c\simeq 
1.35$GeV in our definition Eq. (\ref{DC}). Using this value in 
the amplitude of Eq. (\ref{amp}), we obtain with a 
straightforward computation the branching fraction,
\begin{eqnarray}
   {\cal B}(B\to D_{s1\frac{1}{2}}X) &=&
         \frac{{\cal B}(B\to Xl^+\nu_l)_{{\rm}exp}}{
         \Gamma(B\to Xl^+\nu_l)_{{\rm th}}}\times
          \Gamma(B\to D_{s1\frac{1}{2}}X), 
                    \nonumber \\
          &\simeq&(1.7\times 10^{-2})
          \times(f_A/200{\rm MeV})^2,\label{rate}
\end{eqnarray}
where ${\cal B}(B\to Xl^+\nu_l)_{{\rm exp}}\simeq
0.104$\cite{PDG}, $C_1+\frac{1}{3}C_2\simeq 1.02$\cite{Buras}, 
and the short-distance corrected value of $\Gamma(B\to Xl^+
\nu_l)_{{\rm th}}$ have been used to obtain the numerical 
result. This number is subject to uncertainty of the values 
for $m_b(\simeq 4.5$GeV) and $m_c(\simeq 1.35$GeV). 
The branching fraction is quite large since it is one 
of the dominant factorizable processes.
If we compare two-body decays $B\to D_{s1\frac{1}{2}}
\overline{D}$ and $D_s\overline{D}$, for example, we find
\begin{equation}
      {\cal B}(B\to D_{s1\frac{1}{2}}\overline{D})/
             {\cal B}(B\to D_s\overline{D}) \simeq
              1.5\times (f_A/f_{D_s})^2,
\end{equation} 
where $f_{D_s}$ is the decay constants of $D_s$.

\section{Fragmentation from charm-anticharm jet}
 
 Fragmentation of a heavy meson from a heavy quark was studied 
by many theorists in perturbative pictures\cite{frag}. 
In the heavy limit of a heavy quark, the fragmentation 
functions for $^1P_1$ and $^3P_1$ have a similar 
dependence on $z=2E/m_b$ and ratio of the integrated 
fragmentation probabilities is $\simeq 2/3$ in the perturbative 
calculation\cite{F} when the $^3P_1- ^1P_1$ mass splitting is
ignored. In terms of the ratio of $j=\frac{1}{2}$ to 
$\frac{3}{2}$, this number corresponds to $\simeq 0.88$. After 
the $O(m_s/m_c)$ corrections are included, the ratio shifts a 
little but stays close to unity: $\int D_{j=\frac{1}{2}}(z)dz/
\int D_{j=\frac{3}{2}}(z)dz\simeq 1$. 
Actually, physics of fragmentation of a heavy-light meson is not 
entirely a short-distance process even if one takes the heavy 
quark limit. We do not know of how to estimate nonperturbative 
effects reliably. One can understand complexity of long-distance 
effects if one thinks of cascade feeding from higher resonance 
states.
  
For two $1^+$ states, the orbital wavefunctions are the same 
in nonrelativistic models. Cascade contributions are 
unimportant since higher resonances decay into $DK$ channels. 
If dominant long-distance effects are spin independent like 
the confining force, the {\em ratio} of the fragmentation 
probabilities would not change much with long-distance effects. 
In the absence of a compelling reason for otherwise, 
it is not unreasonable to expect that nonperturbative 
effects do not upset the perturbative prediction 
on the fragmentation ratio:
\begin{equation}
 {\cal B}(e^+e^-\!\to c\overline{c}\to\! D_{s1\frac{3}{2}}X)
 \simeq{\cal B}(e^+e^-\!\to c\overline{c}\to\! D_{s1\frac{1}{2}}X).
\end{equation} 
This is markedly different from $B$ decay in which  
$D_{s1\frac{1}{2}}$ is dominantly produced.

\section{Four-quark state}
  Many theorists proposed\cite{fourquark} that $D_{sJ}(2317)$ 
may be an $0^+$ state of $c\overline{s}q\overline{q}$.
  Although it has been speculated that some of light  
scalar mesons might be four-quark states\cite{Jaffe}, we have 
not yet had a resonance that is proven to be a four-quark meson. 
Consequently, we do not have much knowledge of dynamical 
properties of four-quark states such as production and decay.

We consider four-quark states, which we denote them 
generically by $D_s^{(4)}$. First in $B$ decay. The production 
amplitude for $D_s^{(4)}$ is obtained by superposition of 
a four-quark production amplitude in momentum space:
\begin{equation}
   A_{B\to D_s^{(4)}X}(p_X) =\! 
              \prod_{i=1,2,3}\!\int\frac{d^3{\bf q}_i}{(2\pi)^3}
                 \tilde{\Psi}({\bf q}_i)
             A_{\overline{b}\to (c\overline{s}q\overline{q})
               \overline{c}}(p_X,{\bf q}_i),
                \label{wavefunction}
\end{equation}
where $\tilde{\Psi}({\bf q}_i)$ is the four-quark wavefunction 
in momentum space, ${\bf q}_i$ ($i=1,2,3$) are the relative 
quark momenta inside $D_s^{(4)}$, and $p_X$ is momentum of 
$D_s^{(4)}$. In the loose-binding approximation, the decay 
amplitude turns into a simple form, 
\begin{equation}
     A_{B\to D_s^{(4)}X}(p_X)\simeq \Psi({\bf 0})
   A_{b\to(c\overline{s}q\overline{q})\overline{c}}(p_X,{\bf 0}),
\end{equation}
where $\Psi({\bf 0})$ is the four-quark wavefunction in 
coordinate space with all three relative coordinates set equal 
to zero. $|\Psi({\bf r}_j)|^2$ has dimension of the ninth power 
of energy. For a loosely bound molecular $DK$ state,
\begin{equation}
      |\Psi({\bf 0})|^2=O(\Lambda_{QCD}^6\Delta^3) \label{Psi}
\end{equation} 
where $\Delta$ is the binding energy. For intrinsic four-quark
states in which no $q\overline{q}$ pair is in a color-singlet,
$|\Psi({\bf 0})|^2$ would be comparable with or smaller than 
that of the molecular state since the binding is loose.  
As for the production amplitude, the relevant effective 
interactions are six-quark operators. The dominant interaction 
is of the form 
\begin{equation}
   {\cal L}_{int}\sim (G_{\mu}/\sqrt{2})V_{cb}^*V_{cs}
         (\pi\alpha_s/E^3)
       (\overline{b}\gamma_{\mu}(1-\gamma_5)c)
      (\overline{c}q\overline{q}s)^{\mu},
\end{equation}
where $(\overline{c}q\overline{q}s)^{\mu}$ is a Lorentz vector 
made of four quark fields such as $(\overline{c}\gamma^{\mu}
\partial_{\nu}q)(\overline{q}\gamma^{\nu}s)$, and $E$ is 
determined by the energy scale involved in creation of $q$ 
and $\overline{q}$. Then simple dimension counting gives us 
\begin{equation}
  \Gamma(B\to D_s^{(4)} X)\sim
        G_{\mu}^2|V_{cb}^*V_{cs}|^2(\pi\alpha_s)^2
    \biggl|\frac{\Psi({\bf 0})}{\sqrt{m_{D_s^{(4)}}}}\biggr|^2 
         \frac{m_B^3}{E^6}.  
\end{equation}
Using Eq. (\ref{Psi}), we can estimate the 
branching fraction. If we express in the ratio to the 
branching to $B\to D_sX$ for comparison, 
\begin{equation}
   {\cal B}(B\to D_s^{(4)} X) \simeq 
    (\pi\alpha_s)^2 \biggl(\frac{\Lambda_{QCD}}{E}\biggr)^6
  \biggl(\frac{\Delta^3}{m_{D_s^{(4)}}f_{D_s}^2}\biggr)
  \times {\cal B}(B\to D_s X).   \label{4}
\end{equation}
Since the four-quark binding energy $\Delta$ is much smaller 
than $m_{D_s}^{(4)}$ and $f_{D_s}$, ${\cal B}(B\to 
D_s^{(4)}X)$ is minuscule as compared with
${\cal B}(B\to D_sX)$. The same estimate as Eq. (\ref{4}) 
holds for perturbative fragmentation of $D_s^{(4)}$ from 
$c\overline{c}$ jets. We thus conclude that short-distance 
production of $D_s^{(4)}$ is negligibly small both in $B$ 
decay and in jet fragmentation. In fact, suppression of 
a loosely bound multiparticle state is a general
rule at high energies. The factor $\Delta^3/m_{D_s^{(4)}}
f_{D_s}^2$ in Eq. (\ref{4}) comes from the ratio of the
wavefunctions of $D_s^{(4)}$ and $D_s$. However, there is a 
chance of large long-distance enhancement in the final state 
if $D_s^{(4)}$ and a $c\overline{s}$ meson state of the same 
quantum numbers happen to be almost degenerate in mass. 
In this case the transition of $D_{sJ}(c\overline{s})
\leftrightarrow D_s^{(4)}$ is enhanced by the factor 
$|\Psi({\bf 0})|^2/(M_{D_s^{(4)}}-M_{D_{sJ}})^2$. This 
enhancement factor is square of the $D_{sJ}$-$D_s^{(4)}$ 
mixing itself. We may state therefore that production of 
$D_s^{(4)}$ is possible only if it has a large mixing to 
a quark-antiquark meson state\cite{Pakvasa}. 

How much could $D_s(2317)$ be produced if it is the $0^+$ 
state of $D_s^{(4)}$ mixed with a $c\overline{s}$ state ? 
Since $D_s(c\overline{s})$ of $J^P=0^+$ cannot be produced 
in the factorization, a mixing to it does not help production 
of $D_s^{(4)}$ in $B$ decay. On the other hand $D_s^{(4)}$ 
production through fragmentation would be realized if it 
is mixed substantially with $D_{sJ}(c\overline{s})$ of 
$J^P=0^+$.  Since the ratio of fragmentation 
functions of $0^+$ and $1^+$ is approximately 0.36 for 
$m_s/m_c\simeq 1/3$ in the perturbative calculation\cite{F}, 
even a 50-50 mixing to $D_{s0}(c\overline{s})$ would allow 
fragmentation of $D_s^{(4)}$ only at the level of one fifth 
of $D_{s1\frac{1}{2}}$.  If $D_s^{(4)}$ of 
$J^P=1^+$ mixes with $D_{s1\frac{1}{2}}$ strongly, we may be 
able to see it in $B$ decay. In fragmentation, mixing to 
$1^+$ of either $j=\frac{1}{2}$ or $j=\frac{3}{2}$ helps 
$D_s^{(4)}$ production.
 
\section{Summary}

The BaBar, CLEO, and Belle Collaborations should be able to 
sort out $\Upsilon(4S)$ events and $c\overline{c}$ jet events
by event topology and by $B$ meson decay vertices. 
By analyzing the two types of events separately, we shall 
obtain useful information as to which of $D_{s1}(2460)$ 
and $D_{s1}(2536)$ is $j=\frac{1}{2}$ or $j=\frac{3}{2}$: 
The $1^+$ meson produced abundantly 
in $B$ decay is $D_{s1\frac{1}{2}}$. The $D_{s1\frac{3}{2}}$ 
meson should be looked for in the $c\overline{c}$ jet events. 
If both $D_{s1j}$ states should happen to be produced significantly 
in $B$ decay, or if only one of two $D_{s1j}$ states is produced 
from the $c\overline{c}$ jet, it would be an indication of 
a large mixing between $j=\frac{1}{2}$ and $\frac{3}{2}$, 
that is, failure of the heavy quark approximation to the $c$ 
quark.

\acknowledgements

This work is supported in part by the Director, Office of Science, 
Office of High Energy and Nuclear Physics, Division of High Energy 
Physics, of the U.S.  Department of Energy under Contract 
DE--AC03--76SF00098 and in part by the National Science Foundation
under grant PHY-0098840.

\end{document}